\documentclass[
showpacs,
floatfix,
aps,
prl,
amsmath,
twocolumn,
superscriptaddress,
]{revtex4}

\usepackage{graphicx}
\usepackage{dcolumn}
\usepackage{amsmath}
\usepackage[varg]{txfonts}
\def\eac{\epsilon^{\mbox{\tiny {AC}}}}
\def\edc{\epsilon^{\mbox{\tiny {DC}}}}
\def\pac{\phi^{\mbox{\tiny {AC}}}}
\def\pdc{\phi^{\mbox{\tiny {DC}}}}
\def\oc{\omega_{\mbox{\tiny {C}}}}
\def\rc{R_{\mbox{\tiny {C}}}}

\begin{document}
\title{
Magnetoresistance Oscillations in Two-dimensional Electron Systems Induced by AC and DC Fields
}
\author{W. Zhang}
\affiliation{School of Physics and Astronomy, University of Minnesota, Minneapolis, Minnesota 55455, USA} 
\author{M. A. Zudov}
\email[Corresponding author. Email: ]{zudov@physics.umn.edu}
\affiliation{School of Physics and Astronomy, University of Minnesota, Minneapolis, Minnesota 55455, USA} 
\author{L. N. Pfeiffer}
\author{K. W. West}
\affiliation{Bell Labs, Alcatel-Lucent, Murray Hill, New Jersey 07974, USA}
\received{8 November 2006}

\begin{abstract}
We report on magnetotransport measurements in a high-mobility two-dimentional electron system subject simultaneously to AC (microwave) and DC (Hall) fields.
We find that DC excitation affects microwave photoresistance in a nontrivial way. 
Photoresistance maxima (minima) evolve into minima (maxima) and back, reflecting strong coupling and interplay of AC- and DC-induced effects. 
Most of our observations can be explained in terms of indirect electron transitions using a new, ``combined'' resonant condition.
Observed quenching of microwave-induced zero resistance by a DC field cannot be unambiguously linked to a domain model, at least until a systematic theory treating both excitation types within a single framework is developed.
\end{abstract}
\maketitle

Over the past few years, non-equilibrium transport in very-high Landau levels (LLs) of two-dimensional electron systems (2DES) has become a subject of strong interest.
A large body of theoretical efforts was directed towards microwave-irradiated 2DES, aiming to explain microwave-induced resistance oscillations (MIRO) \citep{zudov:201311,ye:2193}.
On the other hand, it has been known for some time that a phenomenologically similar effect, Hall field-induced resistance oscillations (HIRO) \cite{yang:076801,bykov:245307,zhang:041304} emerges in a 2DES driven by a pure DC-excitation.

Experimentally, microwave (AC)- and DC-induced effects exhibit many similar features.
MIRO appear in microwave photoresistance (MWPR), which is a periodic function of $\eac \equiv \omega/\oc$, where $\omega = 2\pi f$\ and $\oc = eB/m^*$ ($m^*$ is the effective mass) are microwave and cyclotron frequencies, respectively.
HIRO appear in DC-driven 2DES, owing to commensuration of two length scales, cyclotron diameter, $2\rc$ and the real-space separation between Hall field-tilted LLs, $\Delta Y = \hbar \omega_c /eE$ ($E$ is the Hall field).
As a result, the differential resistance, $r\equiv dV/dI$, is periodic in $\edc=\gamma\rc/\Delta Y= \omega_H/\oc$, where $\gamma \simeq 2$ and $\hbar\omega_H = (\gamma \rc) e E$ is the energy associated with the Hall voltage drop across the cyclotron orbit.
Both MIRO and HIRO rely on transitions between high LLs and show similar sensitivity to the sample mobility.
Finally, in the regime of separated LLs and small $\eac$ or $\edc$, strong suppression in resistance can be induced by either AC \citep{willett:026804,dorozhkin:201306,zudov:041303} or DC \citep{zhang:041304,zhang:up} excitation.

On the other hand, our current understanding of these phenomena rests upon different microscopic mechanisms. 
While MIRO were originally discussed in terms of microwave-induced impurity scattering \citep{ryzhii:2078,durst:086803,vavilov:035303}, it is generally believed that dominant contribution comes from inelastic processes leading to a non-trivial distribution function \citep{dorozhkin:577,dmitriev:115316}.
Conversely, HIRO seem to rely on a large-momentum transfer, suggesting importance of short-range disorder, and are believed to stem from intra-LL elastic 
scattering \citep{lei:up,vavilov:up}.

Introducing additional parameters into the AC-driven 2DES can, in principle, help to distinguish between contributions from different mechanisms and even lead to new effects.
In this regard, magnetotransport in 2DES subject to microwave radiation and periodic modulation was recently studied both theoretically \citep{dietel:045329,torres:4029} and experimentally \citep{yuan:075313}.

As far as DC-excitation is concerned, of particular interest is its effect on microwave-induced zero-resistance states (ZRS) \citep{mani:646, zudov:046807}, which are formed in high-quality samples at the MIRO minima.
ZRS are currently understood in terms of absolute negative resistance and its instability, which leads to current domains \citep{andreev:056803,auerbach:196801}. 
Recent experiments with bichromatic microwaves seem to support negative resistance \citep{zudov:236804,durst:752}, but so far there exists no direct experimental evidence for the domain structure.
The theory, however, predicts that domains, and hence ZRS, can only exist below some characteristic current, suggesting DC-excitation as a convenient  probe. 
Despite apparent simplicity of such an experiment, interpretation of the results might be complicated by DC-induced effects, e.g. HIRO, which were not theoretically considered in relation to domains. 

In this Letter we report on magnetotransport studies in a 2DES subject to both microwave and DC excitations.
Remarkably, we observe that 2DES resistance exhibits oscillations governed by a new parameter, $\epsilon=\eac+\edc$ reflecting strong coupling of two excitation types.
Moreover, we can qualitatively explain most of the observations in terms of {\em indirect} electron transitions, viewed as combinations of a vertical jump in energy (microwave absorption) and a sideway jump in space (Zener tunneling) between Hall field-tilted LLs. 
These results are suggestive of a deep relation between AC- and DC-induced effects and   
might provide valuable information about the microscopic origin of the MWPR phenomena, given proper theoretical support.
 
Our Hall bar sample (width $w = 100$ $\mu$m) was fabricated from a symmetrically doped GaAs/Al$_{0.24}$Ga$_{0.76}$As 300 \AA-wide quantum well.
High-quality Ohmic contacts were made by evaporation of Au/Ge/Ni.
After brief illumination, electron mobility, $\mu$, and density, $n_e$ were $\simeq 1.2 \times 10^7$ cm$^2$/Vs and $3.7 \times 10^{11}$ cm$^{-2}$, respectively.
Experiment was performed at a constant coolant temperature $T \simeq 1.5$ K and under continuous microwave radiation. 
While similar results were obtained at other frequencies and intensities, all the data reported here were acquired at frequency $f = 69$ GHz, provided by a Gunn oscillator running at maximum power in tandem with a frequency doubler.
The differential resistance $r$ was measured using quasi-DC (a few Hz) lock-in technique.

Before presenting the experimental data under combined, AC/DC excitation, we briefly review the resonant conditions for MIRO and HIRO. 
MIRO maxima$_{(+)}$ and minima$_{(-)}$ are usually found at $\eac_{n\pm}\simeq n + \pac_{n\pm}$, where $n=1,2,...$ and  $\pac_{n\pm}\simeq\mp\pac_{n}$.
In the regime of overlapped LLs, $\pac_{n}$ is close to $1/4$, but when LLs get separated it becomes progressively smaller with increasing $B$ \citep{zudov:041304,studenikin:245313}.
HIRO, as measured in $r$, obey similar relation,  $\edc_{m\pm} \simeq  m + \pdc_{m\pm}$, with $m = 0, 1, ...$,  $\pdc_{m+} \simeq 0$, $\pdc_{m-} \simeq 1/2$ \citep{zhang:041304}.
With increasing $B$, $\pdc_{0-}$ was observed to decrease \citep{zhang:041304,zhang:up}; in our sample, $\pdc_{0-} \simeq 0.12$ at $B = 1.75$ kG.

\begin{figure}[t]
\resizebox{0.46\textwidth}{!}{
\includegraphics{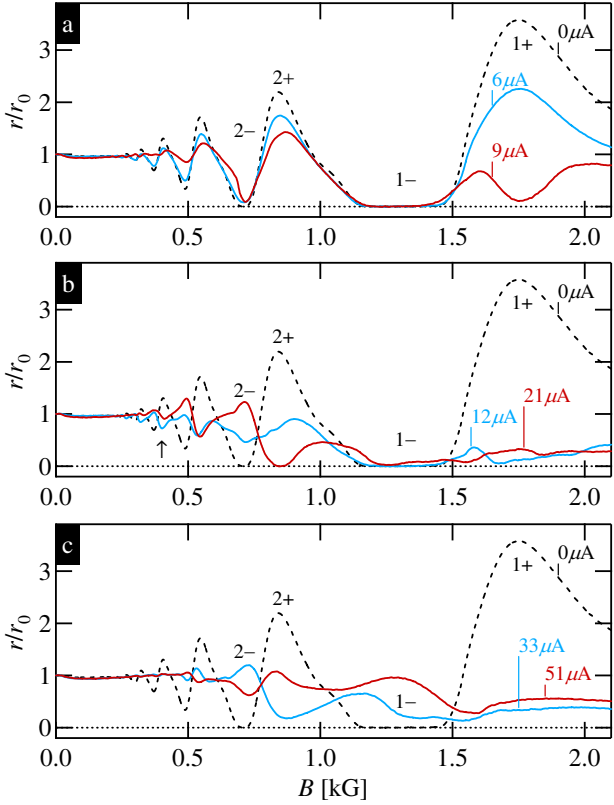}
}
\caption{
Normalized magnetoresistance under microwave illumination of $f = 69$ GHz at $I = 0\,\mu$A (black dashed curve) is plotted in (a), (b), (c), for reference. Differential magnetoresistance is shown in (a) at $6\,\mu$A (light/blue), $9\,\mu$A (dark/red), in (b) at $12\,\mu$A (light/blue), $21\,\mu$A (dark/red), and in (c) at $33\,\mu$A (light/blue), $51\,\mu$A (dark/red). 
}
\vspace{-0.2in}
\label{tr}
\end{figure}
In  Fig.\,\ref{tr} we present typical normalized magnetotransport data, $r/r_0\equiv r(B)/r(0)$, acquired under microwave illumination and different DC excitations.
Zero-bias ($I = 0$) trace (black dashed curve) is included in each panel, for reference.
It shows MIRO and two well-developed ZRS, $\eac_{1-}$ (cf.\,$1-$) and $\eac_{2-}$ (cf.\,$2-$).
In addition,  Fig.\,\ref{tr}(a) shows data acquired at 6 $\mu$A (light/blue) and 9 $\mu$A (dark/red). 
One immediately observes that MWPR is strongly modified even by modest currents.
As far as ZRS are con\-cer\-ned, $\eac_{2-}$ ZRS disappears at 9 $\mu$A (cf.\,$2-$), while $\eac_{1-}$ ZRS remains essentially unaffected (cf.\,$1-$).
Quite surprizingly, fundamental $\eac_{1+}$ peak develops a deep {\em local} minimum (cf.\,$1+$) giving rise to a ``camelback'' structure, while the $\eac_{2+}$ peak (cf.\,$2+$) survives.

In  Fig.\,\ref{tr}(b), we continue with the data taken at higher $I$, 12 $\mu$A (light/blue) and 21 $\mu$A (dark/red).
Examination of the $12 \mu$A trace reveals that high-order MIRO, which were already absent at 9 $\mu$A, reappear, but with the opposite phase (cf.\,$\uparrow$).
At 21 $\mu$A, the ``phase-flip'' progresses into the ZRS regime and 
$\eac_{2-}$ ZRS evolves into a peak (cf.\,$2-$).
Most remarkably, $\eac_{2+}$ peak is transformed into a {\em new} ``ZRS'' (cf.\,$2+$), which coexists with persisting, but about to disappear, $\eac_{1-}$ ZRS (cf.\,$1-$). 
This new ``ZRS'' is a result of a combined, AC/DC excitation.

Data at still higher $I$, 33 $\mu$A (light/blue) and 51 $\mu$A (dark/red), are shown in  Fig.\,\ref{tr}(c).
New ``ZRS'' disappears at $\simeq$ 33 $\mu$A, and evolves back to a peak at 51 $\mu$A (cf.\,$2+$).
The peak also develops at the $\eac_{1-}$ ZRS (cf.\,$1-$).
We thus conclude that DC excitation strongly modifies MWPR and affects not only ZRS, which is expected from the domain model \citep{andreev:056803}, but also the peaks.
In fact, $\eac_{1-}$ ZRS sustain much higher $I$, than the $\eac_{1+}$ peak.

\begin{figure}[t]
\resizebox{0.46\textwidth}{!}{
\includegraphics{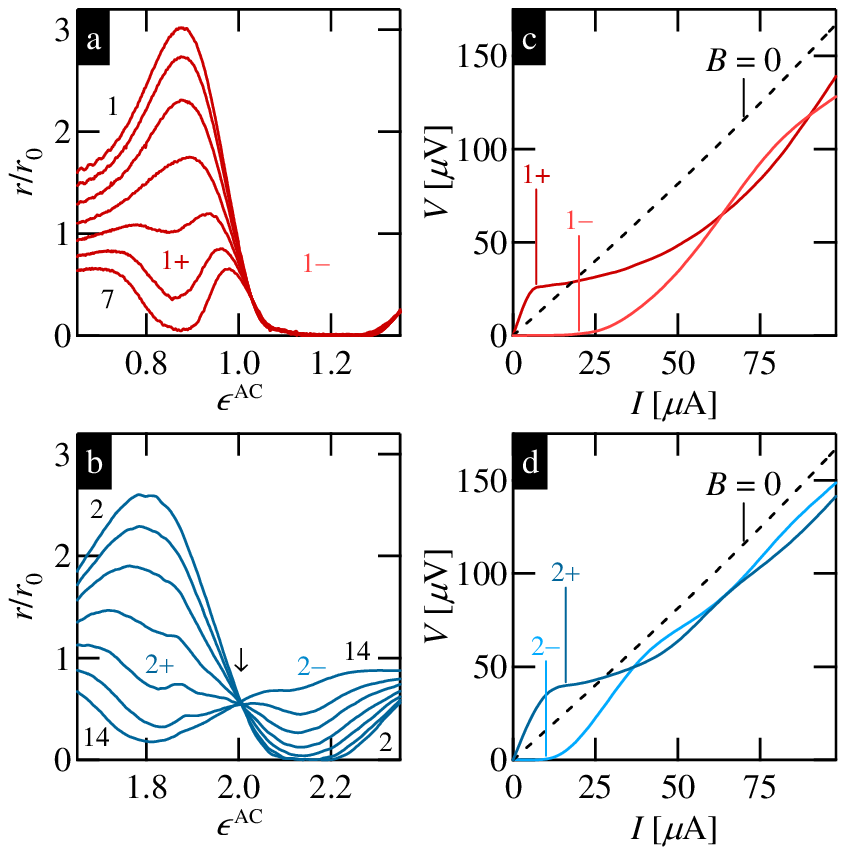}
}
\caption
{
(a) Normalized microwave magnetoresistance about $\eac_{1\pm}$ at $I$ from 1 $\mu$A to 7 $\mu$A, in 1 $\mu$A steps.
(b) Normalized microwave magnetoresistance about $\eac_{2\pm}$ at $I$ from 2 $\mu$A to 14 $\mu$A, in 2 $\mu$A steps.
Both $\eac_{1+}$ and $\eac_{2+}$ peaks show development of a local minimum with increasing $I$.
(c) IVC for $\eac_{1+}$ peak ($1+$) and $\eac_{1-}$ ZRS ($1-$).
(d) IVC for $\eac_{2+}$ peak ($2+$) and $\eac_{2-}$ ZRS ($2-$).
Dashed lines represent IVC at $B = 0$.
}
\vspace{-0.2in}
\label{iv}
\end{figure}
\begin{figure}[t]
\resizebox{0.46\textwidth}{!}{
\includegraphics{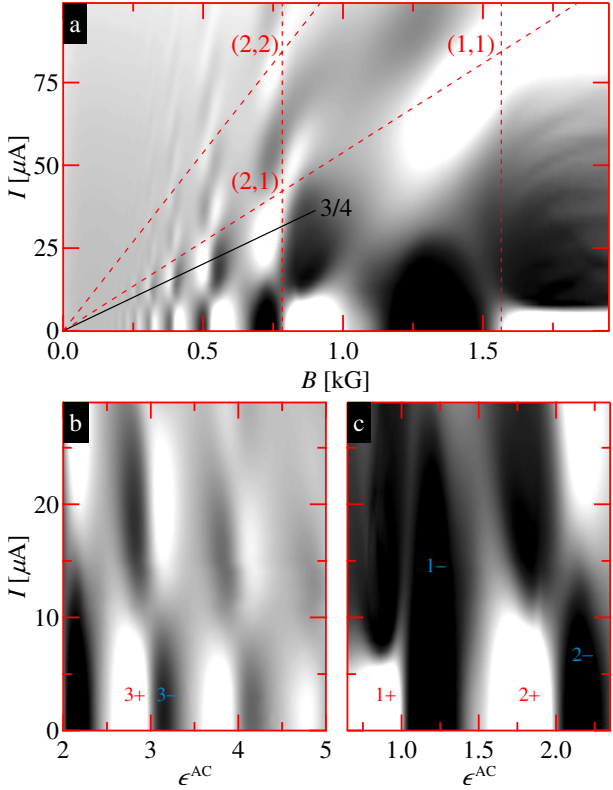}
}
\caption{
(a) Grey scale intensity plot of $r$ in the ($B$, $I$)-plane. 
Light (dark) corresponds to high (low) values of $r$.
Crossings of vertical ($\eac = 1,2$) and inclined ($\edc = 1,2$) dashed lines are marked by ($\eac,\edc$).
Inclined solid line is drawn at $\edc = 3/4$ (see text).
Lower portion of (a) is presented in (b), MIRO regime, and (c), ZRS regime, with $B$ changed to $\eac$. 
}
\vspace{-0.2in}
\label{im}
\end{figure}
\begin{figure}[t]
\resizebox{0.46\textwidth}{!}{
\includegraphics{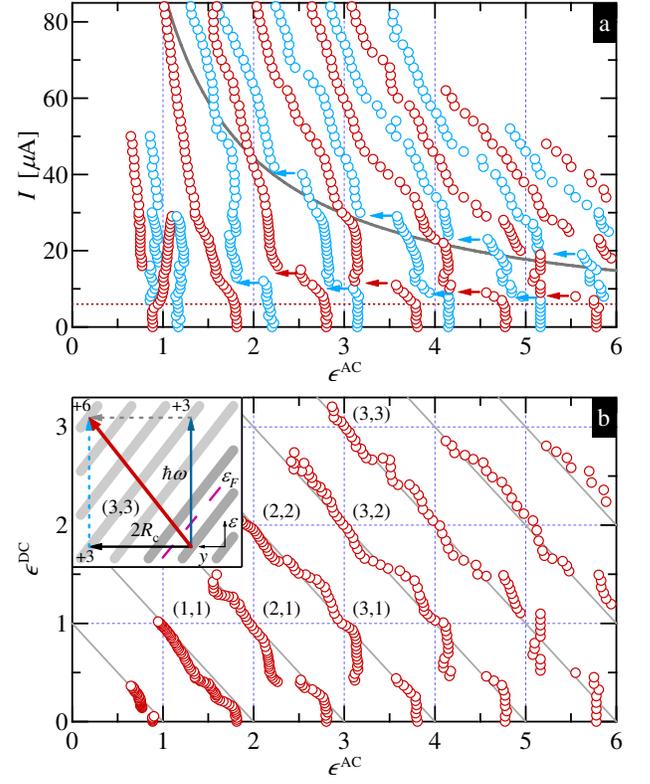}
}
\caption
{
(a) Positions of the $r$ maxima (dark/red) and minima (light/blue) in the ($I$,$B$)-plane.
Solid curve corresponds to $\edc = 1$.
(b) Positions of the $r$ maxima in the ($\eac,\edc$)-plane. 
Solid lines are $\eac + \edc = k$, $k = 1,2,...$ . 
Inset shows LLs (thick lines) at ($\eac,\edc$) = ($3,3$). Horizontal, vertical, and inclined arrows mark electron transitions due to DC, AC, and AC/DC excitations, respectively.
}
\vspace{-0.2in}
\label{fan}
\end{figure}

We now examine the regime of low $I$ where the intriguing ``camelback'' structure was observed at $\eac_{1+}$.
In  Fig.\,\ref{iv}(a) we present microwave magnetoresistance about $\eac_{1\pm}$ under DC current excitation from 1 to 7 $\mu$A, in 1 $\mu$A steps. 
A deep local minimum emerges at the $\eac_{1+}$ peak position (cf.\,$1+$), while the $\eac_{1-}$ ZRS remains essentially unchanged (cf.\,$1-$).
The evolution of MWPR near $\eac_{2\pm}$ is shown in  Fig.\,\ref{iv}(b), where the current was varied from 2 to 14 $\mu$A, in 2 $\mu$A steps. 
A considerably larger $I$ is needed to turn $\eac_{2+}$ peak into a minimum, but one also discerns a weak local minimum at $\eac_{2+}$ (cf.\,$2+$).
At the same time, $\eac_{2-}$ ZRS appears far less persistent (cf.\,$2-$) in comparison with $\eac_{1-}$ ZRS; it disappears at $\simeq$ 6$\mu$A and is already a peak at $\simeq$ 14 $\mu$A. 
In light of dramatic changes seen at MWPR extrema, it is rather strange to observe that zero-response nodes, e.g., $\eac = 2$ (cf.\,$\downarrow$ in  Fig.\,\ref{iv}(b)), remain virtually immune to DC excitation in this range of $I$. 
This observation becomes even more puzzling if one recalls that without microwave radiation $r$ would already be suppressed by about a factor of five \citep{zhang:041304}.

Evolution of the MWPR maxima and disappearance of ZRS can also be examined in current-voltage characteristics (IVC), measured at $\eac_{1\pm}$ and $\eac_{2\pm}$, as shown in  Fig.\,\ref{iv}(c) and (d), respectively.
Indeed, the kink in the IVC corresponding to a deep minimum at $\eac_{1+}$ (cf.\,$1+$) happens much earlier than that at $\eac_{2+}$ (cf.\,$2+$).
This indicates that the strongest, first-order MWPR peak is actually the ``weakest'' with respect to DC excitation.
On the other hand, ZRS disappearance follows the expected trend, with the strongest $\eac_{1-}$ ZRS persisting to much higher $I$ (cf.\,$1-$) than $\eac_{2-}$ ZRS (cf.\,$2-$).
At higher $I$, IVCs, while crossing each other, are roughly parallel to the IVC measured at $B = 0$ (dashed lines).

To further study the effect of DC-excitation on MWPR, we perform $B-$sweeps at different $I$ up to 100 $\mu$A, in 1-2 $\mu$A steps.
The results of these measurements are presented in  Fig.\,\ref{im}(a) as intensity plot in the $(B,I)$-plane, where light (dark) represents high (low) values of $r$.
Periodic patterns of highs and lows are observed if one follows vertical lines (fixed $B$) or some inclines, such as that marked ``3/4''.
Observed periodicity strongly suggests resonant coupling of AC and DC excitations.

In  Fig.\,\ref{im}(b) we focus on the MIRO regime and replot the lower portion of (a) converting $B$ to $\eac$.
For a given $n \gtrsim 3$, the first switch from a max (min) to a min (max) occurs almost at the same $I$ (cf.\,$3+,3-$).
On the other hand, similar image for the ZRS regime, shown in  Fig.\,\ref{im}(c), reveals that there exists a range of $I$ where the switch has already occurred for the maximum, but not for the adjacent minimum.
This is best observed at $\eac_{1\pm}$ (cf.\,$1+,1-$), but is also evident at $\eac_{2\pm}$, further confirming dramatically smaller effect of $I$ on the ZRS than on the peaks.

To examine the resonant condition under both AC and DC excitations, we plot in  Fig.\,\ref{fan}(a) the maxima (dark/red) and minima (light/blue) positions in the $(I,\eac)$-plane.
At small $I$, maxima and minima appear as prescribed by $\eac_{n\pm} = n\mp\pac_{n}$.
With increasing $I$, maxima do not shift up to $I_{n+} \simeq 6 \mu$A, which is roughly independent on $n$ (cf.\,horizontal dotted line). 
Minima, on the other hand, do not change positions up to $I_{n-}$, which grows with decreasing $n$. 
In the ZRS regime, $I_{n-} > I_{n+}$, but in the MIRO regime $I_{n-} \simeq I_{n+}$.
At higher $I$, one observes extrema moving towards lower $\eac$ with jumps across $n$ or $n + 1/2$ (cf.\,$\leftarrow$).
The solid curve is drawn at $\edc = 1$.
It crosses the maxima positions at $\eac = n$ and seems to correlate with the jumps of the minima (cf.\,$\leftarrow$).

We now convert $I$ to $\edc$ according to $\edc=\omega_H/\oc$, with $\omega_H=\gamma \sqrt{2\pi/n_e}I/ew$ and $\gamma=1.9$ \citep{zhang:041304} and replot maxima positions in  Fig.\,\ref{fan}(b).
Remarkably, data roughly fall onto parallel lines, $\eac+\edc\simeq k$, $k = 1,2,...$, apart from deviations associated with the jumps and at very small currents. 
We notice that the deviation is usually small when  $\eac$ and $\edc$ are both integers, as marked by ($n$, $m$) pairs.
This is consistent with the maxima observed in Fig.\,\ref{im}(a) at the crossing points of $\eac = 1,2$ (vertical dashed lines) and $\edc = 1,2$ (inclined dashed lines), as marked by (1,1), (2,1), and (2,2).

As suggested by  Fig.\,\ref{fan}(b), we introduce new parameter $\epsilon = \eac + \edc$, somewhat analogous to that in multi-photon, bichromatic MWPR \citep{zudov:041303}.
The resonant condition under AC/DC excitation takes a remarkably simple form,
\begin{equation} 
\epsilon_{k+} \simeq k ,   \epsilon_{k-} \simeq k + 1/2,
\end{equation}
which prompts to a phenomenological interpretation based on ``combined'' resonances.
As illustrated in the inset of  Fig.\,\ref{fan}(b), the characteristic electron transition can be viewed as a combination of a vertical jump in energy due to microwave absorption owing to AC-excitation and a horizontal jump in space due to elastic scattering by impurities under DC-excitation. 

Derived resonant condition is sufficient to qualitatively understand stronger effect of DC-excitation on MWPR maxima in the ZRS regime.
With increasing $I$ at fixed $\eac = n$ (zero-response nodes of MIRO), the first DC-induced minimum (maximum) should appear at $\edc_{0-} \simeq \pdc_{0-}$ ($\edc_{1+} \simeq 1$).
MWPR maxima (minima), being offset by the phase, will switch to minima (maxima) at $\edc_{+} \simeq \pdc_{0-}+\pac_{n}$ ($\edc_{-} \simeq 1-\pac_{n}$).
In the MIRO regime, $\pac_{n} \simeq 1/4$, $\pdc_{0-} \simeq 1/2$, which gives $\edc_+ \simeq \edc_- \simeq 3/4$.
Indeed, corresponding line, marked ``3/4'' in  Fig.\,\ref{im}(a), passes through the maxima and minima centers.
However, in the ZRS regime, both $\pac_{n}$ and $\pdc_{0-}$ are reduced, yielding $\edc_{+} < \edc_{-}$, in qualitative agreement with the fact that ZRS can sustain higher $I$ than the neighboring peaks. 

In summary, we have studied magnetotransport of a high-quality 2DES in a new experimental regime, i.e., under both microwave and DC excitations.
We have found that DC excitation modifies microwave photoresistance in a non-trivial way, reflecting strong coupling of two excitation types and hinting on a deep relation between AC- and DC-induced effects.
We have observed evolution of the MWPR maxima (minima) into minima (maxima) and back, and concurrent formation of new local minima and new ``ZRS''.
While observed disappearance of ZRS is consistent with the domain picture \citep{andreev:056803}, we emphasize even stronger effect of DC-excitation on the MWPR maxima and that both maxima and minima are described equally well by a new resonant condition.
We can qualitatively understand most of our observations in terms of indirect electron transitions which rely on both inter-LL microwave absorption and inter-LL elastic scattering between Hall field-tilted LLs.
On the other hand, the origin of the deep local minima, immunity of the zero-response nodes, and the existence of domains remains unclear.
Since current understanding of the AC- and DC-induced effects is based on different mechanisms and relies on different types of disorder, it is very desirable to develop a systematic theory which would consider both excitation types simultaneously. 

We thank A. Kamenev, E. Kolomeitsev, B. Shklov\-skii, and I. Dmitriev for discussions and useful comments.
This work was supported by NSF DMR-0548014.
\vspace{-0.25in}

\end{document}